\documentclass[doublecol]{epl2} 
\usepackage{graphicx}
\usepackage{amssymb}
\usepackage[english]{babel}
\usepackage{amsmath}
\usepackage{color}

\newcommand\dd{{\partial }}

\renewcommand{\Pr}{\mathit{Pr}}

\newcommand{\Nu}{\mathit{Nu}}
\newcommand{\Ra}{\mathit{Ra}}

\title{Crossover of the relative heat transport contributions\\ of plume ejecting and impacting zones\\ in turbulent Rayleigh--B{\'e}nard convection}
\shorttitle{Crossover of the relative heat transport contributions  in turbulent RBC} 

\author{Philipp Reiter\inst{1,3} \and Olga Shishkina\inst{1,3} \and Detlef  Lohse\inst{1,2,3} \and Dominik Krug\inst{2,3}}
\shortauthor{Reiter, Shishkina, Lohse, and Krug}

\institute{                    
  \inst{1} Max Planck Institute for Dynamics and Self-Organization, G{\"o}ttingen, DE\\
  \inst{2} Physics of Fluids Group, J.M. Burgers Center for Fluid Dynamics and MESA+ Institute, University of Twente, Enschede, NL\\
  \inst{3} Max Planck UT Center for Complex Fluid Dynamics
}

\abstract{Turbulent thermal convection is characterized by the formation of large-scale structures and strong spatial inhomogeneity. 
This work addresses the relative heat transport contributions of the large-scale plume ejecting versus plume impacting zones in turbulent Rayleigh--B{\'e}nard convection.  
Based on direct numerical simulations of the two dimensional (2-D) problem, we show the existence of a crossover  in the wall heat transport from initially impacting dominated to ultimately ejecting dominated at a Rayleigh number of $\Ra\approx3 \times 10^{11}$.
This is consistent with the trends observed in 3-D convection at lower Ra, and we therefore expect a similar crossover to also occur there. We identify the development of a turbulent mixing zone, connected to thermal plume emission, as the primary mechanism for the crossover. The mixing zone gradually extends vertically and horizontally, therefore becoming more and more dominant for the overall heat transfer.}

\begin{document}

\maketitle

\section{Introduction}
Thermally driven turbulence is omnipresent in nature and technology and its deep fundamental understanding is of utmost relevance for answering various environmental or technological questions. 
As a model system for thermally driven convection, Rayleigh--B\'enard convection (RBC) -- the flow in a box heated from below and cooled from above -- has always been the most paradigmatic and popular one  \cite{ahl09,loh10,chi12}.  It also reflects the intrinsic difficulty of thermally driven flows, namely its  spatially inhomogeneity, including in the lateral direction, due to the formation of large-scale structures.  
Different regions in the flow show different flow features and contribute differently to the overall heat transfer, which is the key global response of  the system to  some given control parameters. In the presence of sidewalls, the spatial inhomogeneity in horizontal direction is obvious. However, due to the formation of large-scale structures, it even holds in the absence of sidewalls, for periodic boundary conditions \cite{poel2015prl,stevens2018,pandey2018,krug2020,blass2021b},  or for very large aspect ratios $\Gamma$ defined as cell width over cell height \cite{emr15}.  
  
The spatial inhomogeneity of the flow must also be reflected in any theoretical approach to understand the heat transfer in RB flow. The simplest example may be the theory of Malkus \cite{mal54}, which assumes a thermal shortcut in the bulk and laminar type heat transport in the thermal boundary layers (BLs), which leads to the scaling law $\Nu \sim \Ra^{1/3} $ for the Nusselt number (the dimensionless heat transfer) as function of the Rayleigh number (the dimensionless temperature difference between top and bottom plate). 
This also holds for the mixing length theory of Castaing \etal \cite{cas89}, the boundary layer theory of Shraiman and Siggia \cite{shr90}, and the unifying theory of Grossmann and Lohse \cite{gro00,gro01,gro04}, which splits the kinetic energy and  thermal dissipation rates into bulk  and boundary-layer/plume contributions, with different scaling behavior. 
Since the kinetic energy and thermal dissipation rates are additive with respect to these contributions and the total dissipation rates can exactly be connected with the overall Nusselt and Rayleigh numbers, this implies that the system response parameters, i.e., the Nusselt and Reynolds numbers,  do not show pure scaling behavior, but a smooth crossover from the dominance of one region to another. 
  
Indeed, different scaling behavior in different regions of the flow were measured in various experiments and direct numerical simulations (DNS).  E.g., for the local heat fluxes, Shang \etal  \cite{sha03,sha04,sha08} measured the scaling close to $\Nu_{loc} \sim \Ra^{1/4}$ near the sidewalls and close to $\sim Ra^{1/2}$ in the bulk. This suggests that for increasing $\Ra$, the latter may take over. 
This is the so-called asymptotic ultimate regime $\Nu\sim  \Ra^{1/2}$,  first suggested by Kraichnan \cite{kra62} and Spiegel \cite{spi71} and indeed found in so-called homogeneous  RB flow \cite{loh03,cal05}, where the flow-driving hot and cold temperature  boundary conditions at the plates have been replaced by a bulk driving with an overall temperature gradient, as well as in radiatively driven convection \cite{lepot2018}. 

Different scaling behavior of the local heat flux 
in different regions of the flow in the lateral direction, reflecting the spatial lateral inhomogeneity of the flow, was also observed in numerical simulations with periodic boundary conditions, despite the periodicity. Van der Poel \etal  \cite{poel2015prl} distinguishes plume ejecting and plume impacting regions, which are seperated by regions dominated by wind-shearing. These regions are set by the  large-scale convection rolls. In general, these large-scale rolls wiggle  laterally due to the turbulent nature of the flow. Therefore, to obtain the local heat flux in the different regions, spatially moving averages have to be performed. In this way, the local dynamics can be disentangled from the slow large-scale movement of the rolls.

 \begin{figure*}\centering
\unitlength1truecm
\begin{picture}(16,6.5)
\put(0.5, -0.2){\includegraphics[height=0.39\textwidth]{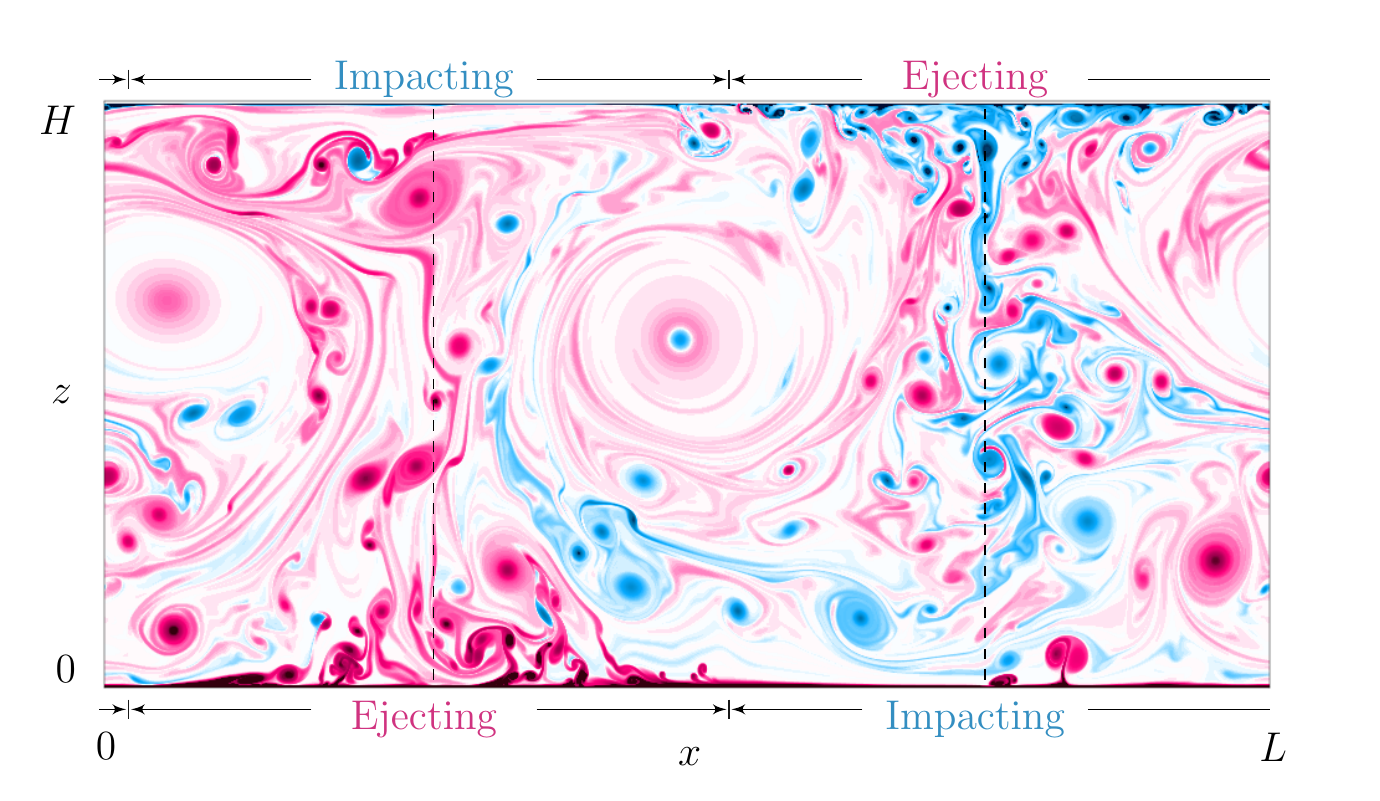}}
\put(13, -0.15){\includegraphics[height=0.39\textwidth]{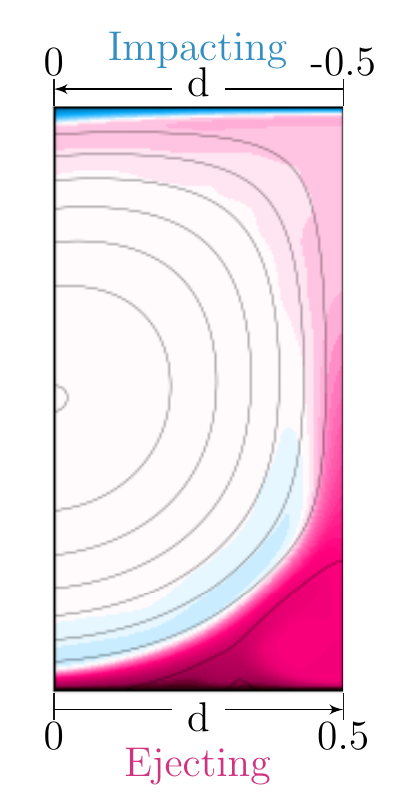}}
\put(0,5.8){$(a)$}
\put(12.5,5.8){$(b)$}
\end{picture}
\caption{(a) A snapshot of the temperature field for $\Ra=10^{11}$. Blue (red) colour corresponds to the temperature below (above) the arithmetic mean of the top and bottom temperatures. The dashed vertical lines show the two locations of the centres of the plume ejecting zones (a hot one on the left and cold a one on the right), which are identified with the  conditional averaging algorithm (see the main text). (b) The mean temperature field together with the mean velocity streamlines, as obtained from the conditional averaging algorithm. Bottom (top) corresponds to the ejecting (impacting) zone. The colour scale is the same as in (a).}
\label{fig.1}
\end{figure*}

Assuming that the large-scale  rolls are of similar size and always located at the same places, for very large $\Ra \ge 10^{13} $ and in two-dimensional (2-D) direct numerical simulations, Zhu \etal  \cite{zhu2018} obtain effective scaling laws: $\Nu_{loc}  \sim Ra^{0.28} $ in the plume impacting region and $\Nu_{loc}  \sim Ra^{0.38} $ in the plume ejecting  region. This implies that for very large $\Ra$, the latter scaling wins for the overall (global) Nusselt number, $Nu \sim Ra^{0.38}$. This is the so-called ultimate regime, which corresponds to $\Nu \sim \Ra^{1/2} $ with logarithmic corrections, as predicted by Kraichnan \cite{kra62} and by Grossmann and Lohse \cite{gro11, gro12}.  
In this regime, the kinetic boundary layer has become turbulent, which is reflected in logarithmic velocity and temperature profiles that allow for an enhanced heat transfer. Note that, without accounting for the movement of the structures in the decomposition, Zhu \etal \cite{zhu2018, zhu2019reply} find that the plume ejecting regions contribute more to the overall heat flux for all $Ra$ explored in their study ($10^{11}\le \Ra \le  4.64 \times 10^{14} $).

In contrast to the findings by Zhu \etal \cite{zhu2018}, for $\Ra$ up to $10^9$ and in three-dimensional (3-D) DNS, Blass \etal  \cite{blass2021b} find that the plume impacting regions contribute more to the overall heat flux. These findings in large periodic boxes ($\Gamma = 32$) are in line with earlier findings in more confined geometries of Reeuwijk \etal~\cite{reeuwijk2008a} ($\Gamma=4$ ) and Wagner \& Shishkina \cite{wagner2012} ($\Gamma=1$). Indeed, such a heat flux distribution would be expected from the fact that the boundary layer thickness grows as the fluid is advected along the plate with correspondingly reduced heat fluxes downstream (i.e. towards the ejecting region). However, Blass {\it et al.} \cite{blass2021b} also find that the dominance of the plume impacting regions diminishes with increasing $\Ra$. Extrapolating their data to higher Ra, they estimate a crossover from impacting to ejecting dominated at $Ra \approx 10^{12}$, which appears consistent with the findings of Zhu \etal~\cite{zhu2018} in 2-D. Blass \etal \cite{blass2021b} use a conditional averaging technique which is superior to a spatially moving average because it allows to extract precise statistics despite movement of the structures or changes in their number and orientation.

 In this paper, we want to reconcile the results of  Zhu \etal \cite{zhu2018}  -- the dominance in the heat flux of the plume ejecting regions in 2-D DNS beyond  $\Ra= 10^{11}$ and of Blass \etal \cite{blass2021b}  -- dominance in the heat flux of the  plume impacting regions in 3-D DNS up to $\Ra= 10^9$, obtained with the conditional averaging technique. We do this by applying the superior conditional averaging technique to the numerical data obtained by Zhu \etal \cite{zhu2018} and to new 2-D direct numerical simulations (DNS) for $\Ra$ down to $10^7$. Our main result is that the observations by Blass \etal \cite{blass2021b} and Zhu \etal \cite{zhu2018} are consistent and robust, and that we can identify the crossover Rayleigh number at which the  heat flux from the plume ejecting regions overtakes that one from the plume impacting regions. This analysis is relevant in the context of the findings of Zhu \etal \cite{zhu2018, zhu2019reply}, who showed that beyond $Ra \ge 10^{13}$, both, the local heat flux in the plume ejecting regions (which grows with increasing $\Ra$) and the overall heat flux, scale steeper than the classical Malkus scaling $\Nu \sim Ra^{1/3}$. This reflects the onset of the ultimate regime around that Rayleigh number, consistent with theoretical predictions \cite{gro00, gro02} and experimental measurements \cite{he12, he12a}.

\section{Numerical Simulations}

In this letter, we restrict us to  2-D DNS of RB flow, with periodic sidewalls and no-slip velocity boundary conditions (BCs). The governing dimensionless Navier--Stokes equations in the Oberbeck--Boussinesq approximations read 
\begin{align}
   \dd{\bf u}/\dd t+{\bf u}\cdot \nabla {\bf u}&=-\nabla {p}+
    \sqrt{Pr/Ra} \nabla^2 {\bf u}+ {\theta}{\bf e}_z \nonumber,\\
    \dd{\theta}/\dd t+{\bf u}\cdot \nabla {\theta}&=1/\sqrt{Pr Ra} \nabla^2 {\theta}, \quad \nabla \cdot {\bf u} =0,
\label{eq:ns}
\end{align}
where ${\bf u}$ is the velocity, $\theta$ the temperature, $p$ the pressure, $t$ the time and ${\bf e}_z$ denotes the unit vector in the vertical direction. The equations have been non-dimensionalised using the free-fall velocity $u_{ff}\equiv (\alpha g \Delta {H})^{1/2}$, the free-fall time $ {H}/u_{ff}$, the temperature difference $\Delta$ between the hot and the cold plate,  and the cell height ${H}$. The parameters $Ra$, $Pr$,  and the aspect ratio $\Gamma$ are 
\begin{align*}
\quad
Ra\equiv \alpha g \Delta {H}^3/(\kappa\nu), \quad Pr\equiv\nu/\kappa=1, \quad
\Gamma\equiv {L}/{H}=2, 
\end{align*}
where $L$ is the lateral extension of the periodic domain, $\alpha$ the fluid thermal expansion coefficient, $\nu$ the viscosity, $\kappa$ the thermal diffusivity and $g$ acceleration due to gravity.

The set of equations (\ref{eq:ns}) is solved numerically using the finite volume code {\sc goldfish} for $\Ra$ from $10^7$ to $3\times10^{12}$.  Complementary, we have reanalysed the flow snapshots from the previously published data series \cite{zhu2018} for $\Ra$ from $10^{11}$ to $10^{14}$, which was generated with the finite-difference solver {\sc AFiD} \cite{ver96,zhu2018afid}. Taken together, the present study covers the parameter range $10^7\leq Ra \leq 10^{14}$. These two computational codes for turbulent RBC were validated against each other, with  excellent agreement \cite{kooij2018}. Besides, we demand the same grid resolution criteria \cite{shi10} in simulations with both codes. Thus, for the overlapping $\Ra$-range, we use grids with the same number of nodes for both codes and with a similar grid nodes clustering near and in the boundary layers attached to the isothermal plates. For further details regarding the computational grids we refer to the supplementary material in Zhu \etal \cite{zhu2018}.

\section{Conditional averaging}

Fig.~\ref{fig.1}(a) gives an impression of the complexity of the flow and its large-scale organization into two counter-rotating circulation rolls, driven by single zones of respectively rising hot plumes and sinking cold plumes. 
We decompose the flow into plume impacting and ejecting zones (the way to do this is explained later), extract statistical information and analyse their individual heat transport contributions. Evidently, this procedure depends on the robustness of the conditional averaging algorithm, which should be applicable in a broad range of $\Ra$ and which, first and foremost, should be able to reliably identify the location of the large-scale rolls.

\begin{figure*}\centering
\unitlength1truecm
\begin{picture}(14,3.8)
\put(0, 0){\includegraphics[width=0.72\textwidth]{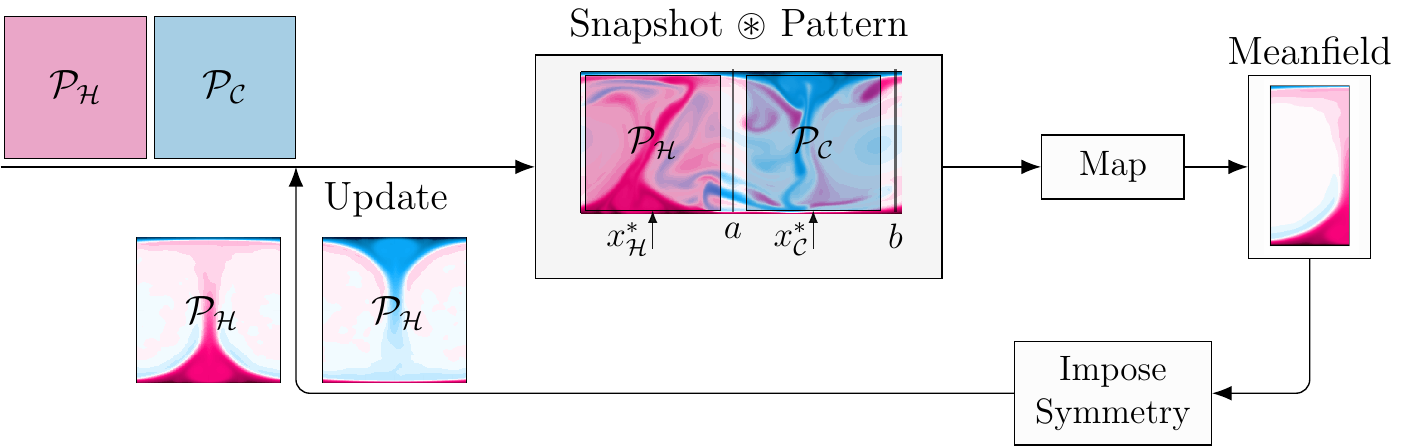}}
\end{picture}
\caption{Schematic overview of the iterative pattern matching algorithm that is used to identify the hot and cold plumes and the conditional average of the fields, based on the horizontal positions of the plumes.}
\label{fig.2a}
\end{figure*}

\begin{figure}
\unitlength1truecm
\begin{picture}(4,5.3)
\put(0.4, -.2){\includegraphics[width=0.44\textwidth]{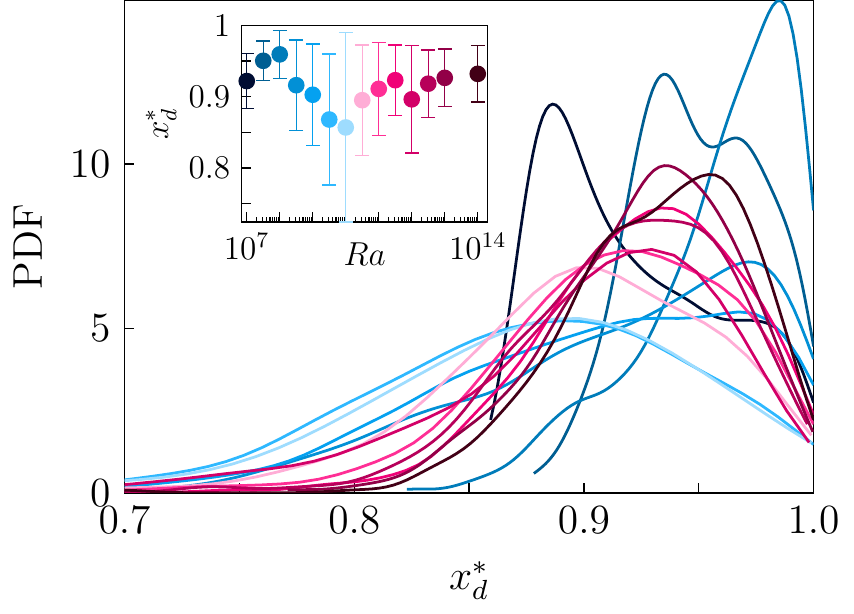}}
\end{picture}
\caption{Probability density distribution of the minimum plume separation $x^*_d \equiv \text{min}|x^*_{H}-x^*_{C}|$. The different curves represent different $\Ra$.  The inset figure shows the mean and one standard deviation of the minimal distance between the hot and cold plumes, for different $\Ra$. The colour scale ranges from blue (smaller $\Ra$) to red (larger $\Ra$), according to the inset figure.}
\label{fig.2b}
\end{figure}

In a 3-D domain, the first choice for the large-scale roll identification would be the technique of Krug \etal  \cite{berghout2021,krug2020,blass2021b}, which was developed to identify 3-D superstructures in turbulent RBC. 
The method is based on the observation that there exists a pronounced scale separation between the turbulent thermal superstructures and small-scale turbulent fluctuations so that, after applying a low-pass filter at an intermediate wave-number ($k\approx2/H$), what remains is a visually convincing representation of the large-scale structures. However, the turbulence cascades (and hence the spectral energy distributions) are different in 2-D and 3-D flows. As a result, the 2-D flows studied here lack the needed distinct scale separation, which impedes the applicability of the technique \cite{krug2020,blass2021b} to 2-D RBC.

On the other hand, one of the advantages of the 2-D confined ($\Gamma=2$) cell is that the configuration of the large-scale circulation (LSC) is stable, which means that one can safely assume the number of the convection rolls to be fixed for all times. 
This is beneficial in so far, that the problem reduces to finding the horizontal positions of the hot and cold large-scale plumes as functions of time, i.e. $x^*_\mathcal{H}(t)$ and $x^*_\mathcal{C}(t)$. These positions are indicated by the dashed lines in Fig.~\ref{fig.1}(a). 
The functions $x^*_\mathcal{H}(t)$ and $x^*_\mathcal{C}(t)$ are generally independent so that, in particular, the distance between the plumes changes with time.

To find $x^*_\mathcal{H}(t)$ and $x^*_\mathcal{C}(t)$, we employ a pattern (or, equivalently, template) matching algorithm. The idea is simple. 
We check the flow fields for the presence of some pattern by moving a template (horizontally) over the flow field and measure their similarity via convolution. 
In doing so, the templates $\mathcal{P_{H}}$ and $\mathcal{P_{C}}$ are chosen such that they resemble the structure of the region of interest, i.e. the hot ($\mathcal{H}$) upwelling and cold ($\mathcal{C}$) downwelling plumes.
The (circular) convolution can be formally defined as 

\begin{equation}
    \left(\mathcal{P_{H(C)}}\circledast\theta\right)\left(x^\prime\right) = \int_0^L\int_0^H\mathcal{P_{H(C)}}(x^\prime-x,z)\theta(x,z)dxdz,\\
\label{eq.2}
\end{equation}
and the position of the maximal correlation gives the location of the corresponding hot or cold, plume, i.e.

\begin{equation}
x^*_\mathcal{H(C)} = \mathrm{max}\left( \mathcal{P_{H(C)}}\circledast\theta \right).
\label{eq.3}
\end{equation}

This approach is similar to the one proposed by Kooloth \etal  \cite{kooloth2021} with the difference, that their templates are given by previously computed steady state solutions. In contrast, our templates (patterns) are generated iteratively. 

The entire process is schematically depicted in Fig.~\ref{fig.2a}. 
We begin by choosing initial templates $\mathcal{P_{H(C)}}$: functions, which are independent of the vertical coordinate and vary as $\cos(x)$ in the horizontal direction.
Then we calculate $x^*_\mathcal{H}$ and $x^*_\mathcal{C}$, according to Eqs. (\ref{eq.2}) and (\ref{eq.3}). 
Once $x^*_\mathcal{H}$ and $x^*_\mathcal{C}$ are found, we construct the conditional mean field of the hot plume by mapping the variable width subdomains $[b, x^*_\mathcal{H}]$ and $[a, x^*_\mathcal{H}]$ -- assuming reflection symmetry -- onto the fixed interval $[0,0.5]$. Here, $a$ and $b$ denote locations, halfway between $x^*_\mathcal{H}$ and $x^*_\mathcal{C}$, as sketched in  Fig.~\ref{fig.2a}. A similar procedure is applied for the region around $x^*_\mathcal{C}$, which gives a second field, i.e., the conditional field of the cold plume. Afterwards, we merge both fields by using the statistical symmetry of the hot and cold plumes. In this way, all snapshots are processed and successively added to the conditional time averaged mean field, as it is shown in Fig.~\ref{fig.1}(b). 
Finally, from this mean field we can generate new templates $\mathcal{P_{H(C)}}$, by reapplying the inherent symmetries and then restart the algorithm. In practice, the algorithm converges after one iteration and delivers convincing and robust positions of $x^*_\mathcal{H}$ and $x^*_\mathcal{C}$. 
We confirmed that stretching and squeezing of the flow fields while mapping does not distort the global response characteristics like $\Nu$, which remained practically unaffected ($<1\%$ deviation) by these manipulations.

As noted, the templates are first initialised with vertically independent cosine functions. Therefore, the first step of the algorithm is equivalent to a cosine fit method, which is often used to identify the LSC in cylindrical cells \cite{cio97, brown2006}. This method, however, inherently constrains the size of the LSC and fixes the relative size of the large-scale hot and cold plumes and the distance between them. 
To get an impression about the limitations of the cosine fit method, we evaluate the probability density function (PDF) and the mean and standard deviation of the (minimal) relative distance between $x^*_\mathcal{H}$ and $x^*_\mathcal{C}$, or, in simple terms, how close do the hot and cold plumes approach each other (see Fig.~\ref{fig.2b}). 
We find that for all $Ra$ this distance varies quite substantially within the range between 0.8 to 1, which shows that the relative motion of the plumes is significant. As a consequence, the cosine fit method leads to rather ``blurry'' looking mean fields, while the mean fields from our conditional averaging algorithm appear more ``in-focus''.  

\section{Results}

In the following analysis, we will make use of two Nusselt number definitions. The first one is based on the global heat transport and defined as
\begin{equation}
    \Nu = \sqrt{\Ra\,\Pr} \langle\overline{u_z\theta}\rangle_V - \langle\partial_z \overline{\theta}\rangle_V,
\end{equation}
where the overline represents the conditional time average and $\langle \cdot \rangle_V$ denotes a volume average. The second one, given by
\begin{equation}
   \overline{Nu}(d) =  -\partial_z \overline{\theta}|_{z=0\text{ ( or } z=H \,)}
\end{equation}
characterizes the local wall heat transport as a function of the conditioned horizontal coordinate $d$. As depicted in Fig.~\ref{fig.1}(b), $d$ is defined such that the ranges $-0.5\le d < 0$ and $0< d \le 0.5$ correspond to plume ejecting and impacting regions, respectively.

\begin{figure}
\unitlength1truecm
\begin{picture}(4,6.4)
\put(0.5, 1.0){\includegraphics[width=0.4\textwidth]{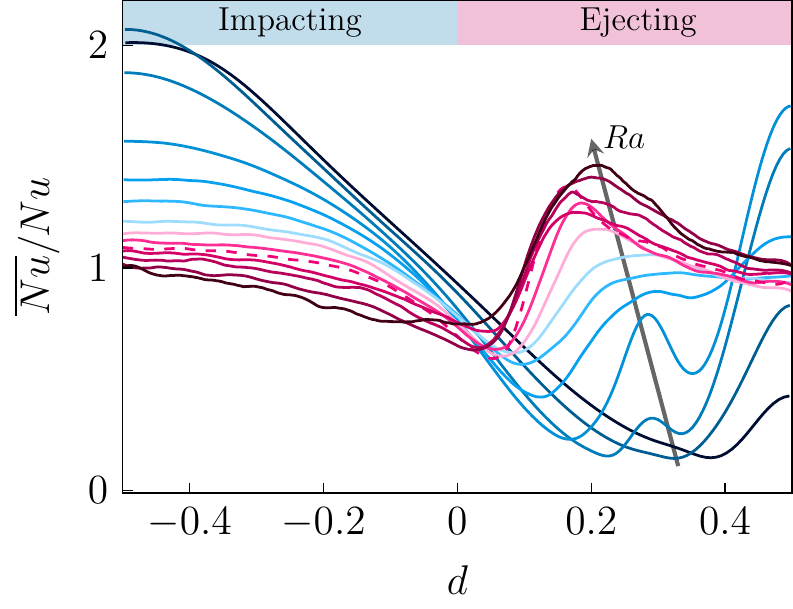}}
\put(1.26, -0.4){\includegraphics[width=0.382\textwidth]{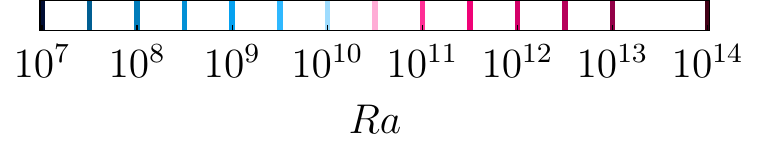}}
\end{picture}
\caption{Horizontal distribution of the normalized heat flux at the plates for small (blue) to large (red) $Ra$. Dashed line: $\Ra=3\times10^{11}$. 
}
\label{fig.3}
\end{figure}

We start off by analysing the horizontal distribution of $\overline{\Nu}$ normalized by $Nu$. Generally, the thermal BL grows as the fluid travels downstream along the plate, i.e. when proceeding from the large scale impacting to the emitting region. Correspondingly, the local heat transfer in the laminar and weakly chaotic regime is expected to decrease along this direction \cite{bej93}. 
This is seen in \cite{reeuwijk2008a, wagner2012,blass2021b} and consistently also here (Fig.~\ref{fig.3}) at small $\Ra$, for which the heat transport reaches its maximum in the impacting zone and then gradually decreases with increasing $d$ in the ejecting zone. However, as $\Ra$ increases, we observe two distinct departures from this picture, which both lead to heat transport enhancements in the ejecting zone.

The first enhancement occurs in the center of the ejecting zone, at $d>0.35$, where a significant peak in $\overline{\Nu}/\Nu$ emerges initially as $Ra$ is increased beyond $10^7$. However, this peak reaches a maximum at $Ra = 10^9$ and eventually subsides again with even stronger thermal driving.
Based on an analysis of instantaneous flow fields, we identified the formation of small recirculation regions, which lead to secondary circulation cells, as the source of this behaviour.  These recirculations disappear for $\Ra\gtrapprox10^{10}$ and therefore do not play a role beyond. Besides that, this phenomenon is likely (and certainly in it's strength) peculiar to 2-D RBC, since it is not observed in comparable 3-D studies \cite{reeuwijk2008a,wagner2012, blass2021b}.

Of more general relevance is the peak that first emerges at $\Ra\approx 10^8$ in the region $0.2< d < 0.35$ (see Fig.~\ref{fig.3}) and which ends up dominating the overall wall heat transport for larger $\Ra$. From Fig.~\ref{fig.1}(b), we can verify that this part of the domain, the "leg" of the large-scale plume, is the predominant origin of small-scale turbulent plumes that emit from the BLs. 
These turbulent plumes are able to effectively mix their surroundings, thus inducing an increase of the vertical and horizontal heat transport. 
Hence, this part of the domain can be seen as the turbulent mixing zone, as suggested by Castaing \etal  \cite{cas89}. We further note that a similar increase of the heat transfer in streamwise direction is also observed in connection with the shear-driven laminar-turbulent transition of the BL \cite{wu2010}. 
The peak first occurs at $d \approx 0.35$ at $Ra = 10^8$ but its location gradually shifts towards lower values of $d$ as $Ra$ is increased. This lends some support to the hypothesis of van der Poel \etal \cite{poel2015prl}, who surmised that the transition to ultimate scaling is driven by a spreading by the plume-ejection dominated region. However, especially at the highest $Ra$ studied here, the simultaneous growth in peak magnitude with increasing $Ra$ appears to be an even more relevant factor in shifting the balance in the heat transfer distribution towards the ejecting side.

\begin{figure}
\unitlength1truecm
\begin{picture}(4,6.2)
\put(.8,6.3){$(a)$}
\put(4.65,6.3){$(b)$}
\put(.8,3.3){$(c)$}
\put(4.65,3.3){$(d)$}
\put(.0, 3.5){\includegraphics[width=0.26\textwidth]{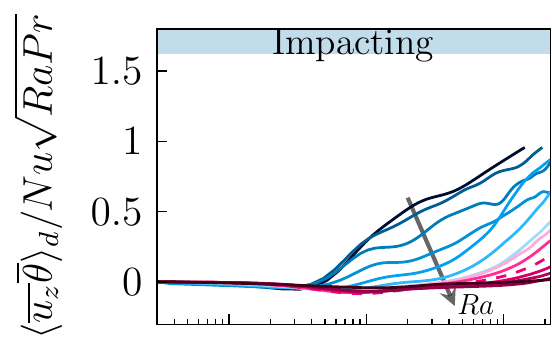}}
\put(4.8, 3.5){\includegraphics[width=0.199\textwidth]{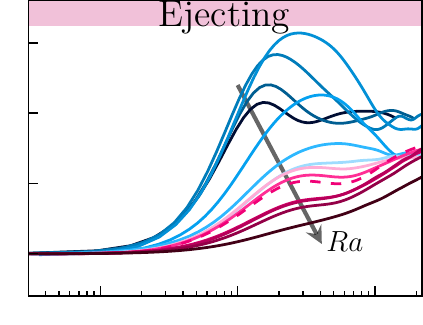}}
\put(.0, -0.4){\includegraphics[width=0.26\textwidth]{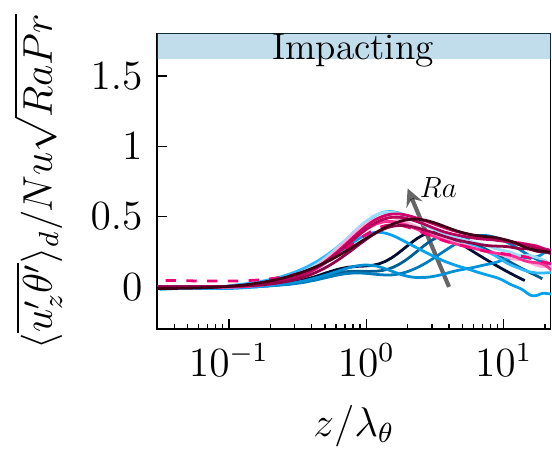}}
\put(4.8, -0.4){\includegraphics[width=0.199\textwidth]{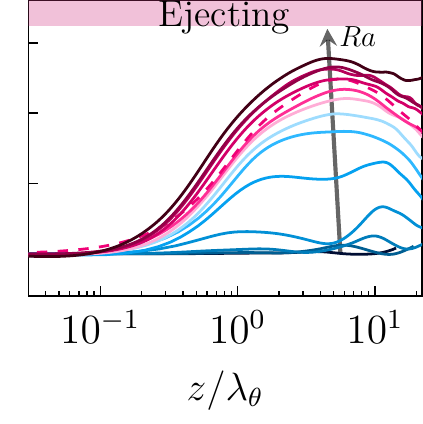}}
\end{picture}
\caption{Vertical distribution of the horizontally averaged mean (top) and turbulent (bottom) convective heat transport in the impacting (a,c) and ejecting (b,d) zone  for small (blue) to large (red) $\Ra$. Dashed line: $\Ra=3\times10^{11}$.
}
\label{fig.4}
\end{figure}
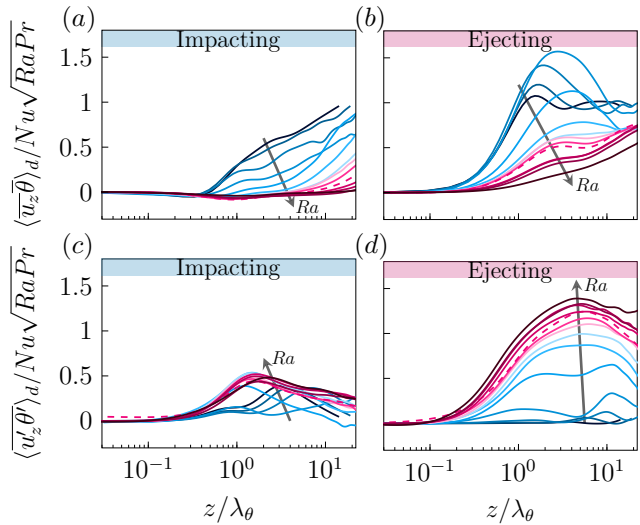

To shed more light on the local heat transport mechanisms, we proceed and decompose the convective heat transport $\overline{u_z\theta}$ into its mean field ($\overline{u}_z\overline{\theta}$) and turbulent contributions ($\overline{u_z^\prime\theta^\prime}$) according to
\begin{equation}
   \langle\overline{u_z\theta}\rangle_d = \langle\overline{u}_z\overline{\theta}\rangle_d + \langle\overline{u_z^\prime\theta^\prime}\rangle_d,
\end{equation}
where $\langle \cdot \rangle_d$ denotes a horizontal average taken either across the impacting ($d<0$) or the ejecting ($d>0$) region. The mean and turbulent convective heat transport profiles in vertical direction, normalized with the mean thickness of the thermal BL $\lambda_\theta=1/(2\Nu)$, are shown in Fig.~\ref{fig.4}. 

Figures \ref{fig.4} (a,~b) show that for small $\Ra$, the mean field transport is dominant in the convective heat transfer for both, impacting and ejecting zones. However, its relative contribution weakens with increasing thermal driving such that the large-scale circulation ultimately plays no significant role in the convective heat transport at high $Ra$. This behaviour is especially apparent in the impacting zone Fig.~\ref{fig.4} (a), where the mean convective transport vanishes almost completely, whereas in the ejecting zone (Fig.~\ref{fig.4}~b), the mean convective transport remains of significant importance. Moreover, the contribution in the ejecting zone first rises before it starts to decay. This is another manifestation of the recirculation regions mentioned earlier, and therefore it is no coincidence that also here the strongest effect is observed at $\Ra=10^9$.

We turn the focus now to the turbulent convective heat transport shown in Figures \ref{fig.4} (c,~d). As expected, turbulent mixing becomes the predominant heat transport mechanism at large $\Ra$, but again, the impacting and ejecting zones behave characteristically differently.
In the impacting zone (Fig.~\ref{fig.4}~c), the turbulent transport contribution initially increases with increasing $\Ra$ before it saturates. Above $\Ra\approx10^{10}$, the curves almost collapse onto a single curve, which was observed similarly in Blass \etal \cite{blass2021b}. This indicates that above a certain $\Ra$, the relative contribution of the turbulent heat transport in the impacting region does not increase significantly anymore. 
In the ejecting region ($0 \le d\le 0.5$), however, the relative importance of turbulent transport increases gradually with increasing $\Ra$, confirming again the trends observed earlier in 3-D convection \cite{blass2021b}. Evidently, the aforementioned ejecting plumes create an efficient turbulent mixing zone which becomes more and more important and ultimately dominates the heat transport mechanisms. Moreover, the mixing zone increases in size with increasing $\Ra$, compared to the thermal BL thickness, and reaches its maximal effectiveness at $z\approx 5\lambda_\theta$. 
This complements the observations by Schumacher \cite{schumacher2008}, who showed that the extension of this mixing zone can be significantly larger than the thermal BL. Surprisingly, the location of this maximum is relatively robust with respect to  changes in $\Ra$. It is further noteworthy that turbulent transport is dominant even deep inside the BL ($z<\lambda_\theta$) once $Ra \gtrapprox 10^{10}$.

\begin{figure}
\unitlength1truecm
\begin{picture}(4,9.7)
\put(0.0,9.2){$(a)$}
\put(0.0,5.7){$(b)$}
\put(0.1, 5.7){\includegraphics[width=0.44\textwidth]{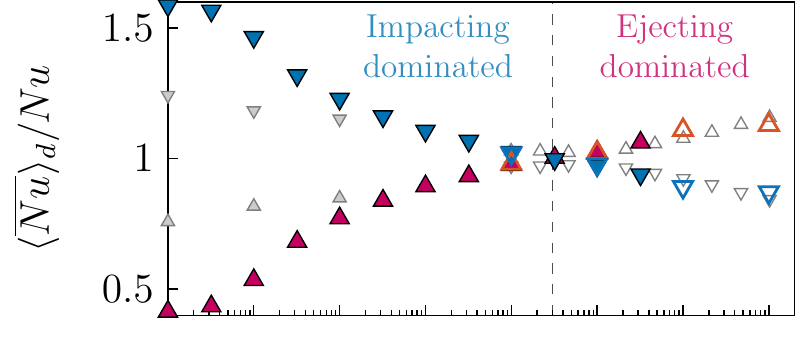}}
\put(0.0, -0.2){\includegraphics[width=0.46\textwidth]{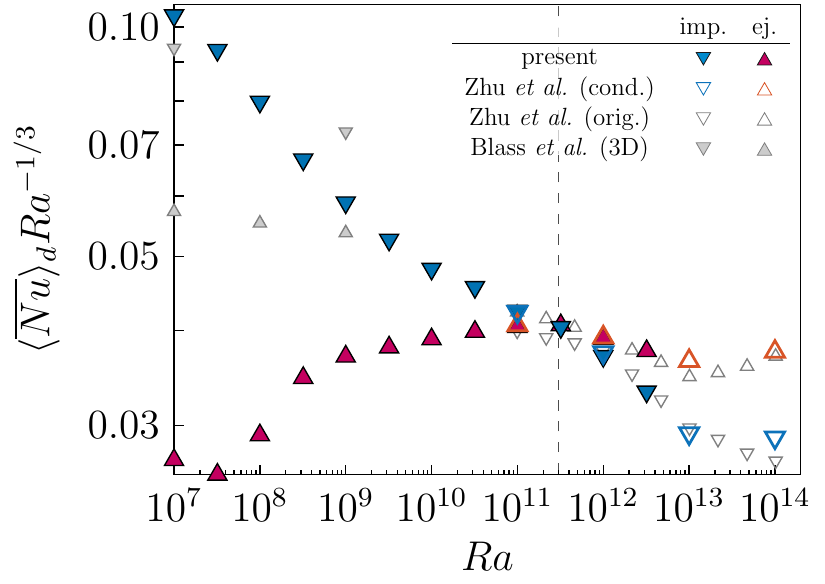}}
\end{picture}
\caption{Local wall heat flux $\overline{Nu}$ compensated by $(a)$ the global Nusselt number $\Nu$ and $(b)$ $\Ra^{-1/3}$. 
Results from the new simulations (closed colour symbols) and results obtained from the analysis of the flow snapshots from Zhu \etal \cite{zhu2018} (open colour symbols) show  the Nusselt numbers evaluated in the plume impacting regions (blue downwards triangles) and in the plume ejecting regions (red upwards triangles). 
For comparison, we show the results for the 2-D case, as reported in Zhu \etal  \cite{zhu2018} (closed grey symbols), and the results for the 3-D case, as reported in  
Blass \etal  \cite{blass2021b} (open grey symbols). 
A crossover of the heat transport contributions to the total heat flux from the plume impacting regions and from the plume ejecting regions occurs at $\Ra \approx 3 \times 10^{11}$.}
\label{fig.5}
\end{figure}

As mentioned in the introduction, for low and intermediate $\Ra$ in 3-D RBC, the impacting zones dominate the wall heat transport \cite{blass2021b}, whereas for large $\Ra$ in 2-D RBC, the ejecting zones were found to contribute the majority of the heat transport \cite{zhu2018}. Yet a direct link between these observations has up to now been missing. With this in mind, we compare contributions of the emitting and ejecting regions to the total wall heat transport in Fig.~\ref{fig.5}.
Thanks to the wide range of $Ra$ available here, we now observe a clear crossover of the contributions from the ejecting and impacting regions at $\Ra\approx 3\times10^{11}$. At this critical Rayleigh number (highlighted as dashed lines in Fig.~\ref{fig.3} and ~\ref{fig.4}), the dominance of the contribution from the impacting zone changes to the dominance of the contribution from the ejecting zone.
The data by Zhu \etal \cite{zhu2018} also show such a crossover, if we apply the conditional averaging as proposed in the present study. Therefore, the dynamic tracking of the LSC is the key to a successful individual statistical description of the different zones. 
Additionally, from Fig.~\ref{fig.5}(a), we find that the 3-D and 2-D cases show increasingly similar behaviour as $\Ra$ increases. This gives confidence that the observed trends in the $Nu$ distribution are indeed driven by the increase in $Ra$ and not predominantly related to differences between 2-D and 3-D flows. We would expect a similar crossover to occur in 3-D as well at sufficiently high $Ra$.


\section{Conclusion}
By means of direct numerical simulations and using a conditional averaging technique we explored the properties of the plume impacting and plume ejecting zones in horizontally periodic 2-D RBC. This study covers the range $10^7 \le \Ra \le 10^{14}$, thus bridging the $Ra$ gap between the corresponding studies of Blass \etal \cite{blass2021b} (3-D) and Zhu  \etal  \cite{zhu2018} (2-D). We provide an unifying picture of the relative heat transport importance of ejecting and impacting zones across $Ra$ and show the existence of a crossover from an impacting dominated to an ejecting dominated local wall heat transfer at $Ra\approx3\times10^{11}$. This trend is connected to an increase in the convective heat transport at the leg of the large-scale plume. 
Specifically, we identify the turbulent convective transfer to become the dominant transport mechanism, which is reflected in a gradual growth with $Ra$ of the turbulent convective heat flux $\overline{u_z^\prime\theta^\prime}$.
The turbulent mixing zone reaches its peak efficiency at a vertical distance of about five thermal boundary layer thicknesses from the plate and it gradually expands in size with increasing $Ra$, thus occupying an ever larger fraction of the domain. 
It remains to be verified whether such a crossover towards the dominance in the heat transport of the thermal plume ejecting regions also exists in 3-D turbulent convection, but our results strongly suggest so.


\acknowledgments
We  thank R.~Verzicco, R.~J.~A.~M.~Stevens, X.~Zhu, and A.~Blass for the ongoing collaboration and for discussions and the latter two also for providing us with the  data. 
We acknowledge the Twente Max-Planck Center, the support by the Deutsche Forschungsgemeinschaft (DFG) under the grant Sh405/10 and SPP 1881 "Turbulent Superstructures", and the Leibniz Supercomputing Centre (LRZ). We also thank the Max-Planck HPC Teams in G{\"o}ttingen and Munich for their generous technical support and additional computational resources.

\bibliographystyle{eplbib}
\bibliography{literatur} 


\end{document}